\documentclass[10pt,a4paper]{article}
	\usepackage[utf8]{inputenc}
	\usepackage[T1]{fontenc}
	\usepackage{amsmath}
	\usepackage{amsfonts}
	\usepackage{cite}
	\usepackage{amssymb}
	\usepackage{graphicx}

	\author{Prashant M. Gade$^{1,*}$ and Sachin B. Bhalekar$^{2,3}$\\
	$^1$ Department of Physics,\\ Rashtrasant Tukadoji Maharaj Nagpur University, Nagpur, India\\
	Email: prashant.m.gade@gmail.com (*Corresponding author)\\
	$^2$Department of Mathematics, \\Shivaji University, Kolhapur, India\\
	$^3$School of Mathematics and Statistics, \\University of Hyderabad, Hyderabad, India\\
	Email: sachin.math@yahoo.co.in
	}
	\title{
	On fractional-order maps and their synchronization
	}

\begin{document}
	\maketitle
	%On fractional-order maps and their synchronization.
	\begin{abstract}
	We study the stability of linear fractional order maps. We show that
	in the stable region, the evolution is described by Mittag-Leffler
	functions and a well defined effective Lyapunov exponent can be
	obtained in these cases. For one-dimensional systems, this exponent
	can be related to the corresponding fractional differential equation.
	A fractional equivalent of map $f(x)=ax$ is stable for
	$a_c(\alpha)<a<1$ where $\alpha$ is a fractional order parameter and $a_c(\alpha)\approx -\alpha$.
	For coupled linear fractional maps,
	we can obtain `normal modes' and reduce the evolution to
	effectively one-dimensional system.
	If the eigenvalues are real the stability of the coupled system
is dictated by the stability of effectively one-dimensional normal modes.
	For complex eigenvalues, we obtain a much richer picture. However,
	in the stable region, the evolution of modulus is dictated by
	Mittag-Leffler function and the effective Lyapunov exponent is
	determined by modulus of eigenvalues.
	We extend these studies to synchronized fixed points of fractional nonlinear maps.
	\end{abstract}

	\section{Introduction}

	After May's influential paper\cite{may}, the difference equations
	gained wide popularity in all walks of science. They
	were studied from
	the viewpoint of nonlinear dynamics and chaos which was ubiquitous
	in disciplines ranging from ecology to economics. Mathematicians
	studied it as an interesting object in its right\cite{strogatz}.
	The studies on maps are computationally less intensive. In certain
	cases, it was easier to track them analytically. Feigenbaum's study
	of period-doubling bifurcations in logistic maps is an
	example\cite{feigenbaum1,feigenbaum2}.
	Lower order difference equations can show the same phenomena
as higher-order differential equations. For example, we can observe chaos in
	a logistic map that has a single variable while we need at least three first-order autonomous differential equations to observe chaos.
	Insights
	gained from studies in difference equations were often
	(though not always) useful in studies in differential equations
	and vice versa. From control problems to synchronization schemes,
	many theoretical ideas have found applications in both difference
	and differential equations\cite{ott}. In this work,
	we study fractional linear difference equations and coupled fractional linear difference equations. We show that the
	analysis of linear fractional
	difference equations has remarkable similarities with corresponding
	fractional differential equations.
	For coupled difference equations, the Jacobian, its eigenvalues
	and eigenvectors play a central role. We show that similar concepts
	can be very useful for coupled fractional difference equations.

	Synchronization of dynamical systems has been extensively studied in the past three
	decades both theoretically as well as experimentally. Exact synchronization
	for master-slave type systems as well as synchronization for mutually coupled
	systems has been investigated extensively. Apart from exact synchronization,
	several other types such as anti-synchronization, anticipated synchronization,
	lag synchronization, phase synchronization, generalized synchronization, etc. have been studied. For spatially extended systems, the connectivity matrix of underlying topology plays an important role in synchronization.
	For exact synchronization, we have a clear mathematical formalism for finding necessary and sufficient conditions for synchronization.

	Ordinary differential equations have successfully described a variety of
	physical system and found countless applications since Newton and
	Leibniz introduced them. However, memory plays an important role in many physical systems. Apart from a mathematical curiosity, such systems are
	modeled by fractional-order differential equations from viewpoint
of applications.
	For fractional-order systems, fractional order maps have been
	introduced recently. These systems are not very well investigated yet.
	In a previous work\cite{pakhare},
	a coupled map lattice model of fractional order
	maps was studied. It was found that it is possible to have synchronization even in the thermodynamic limit in these systems. However,
	the error reduced as a power-law. The exponent
is the same as the fractional order of maps. We give certain
	pointers to understand these findings analytically in this work.

	As in differential equations, we start by studying the stability
	of a fixed point in linear systems.
	We study two coupled fractional maps
	and derive conditions for
	synchronization analytically.
	For simple linear maps the
	bounds of synchronization and its relation to
	Mittag-Leffler function can be shown analytically.
	Asymptotically, Mittag-Leffler function behaves as a power-law
	and this could be the reason for the observation
	in previous work.
	For a linear function with constant
	slope, this relation can be shown very clearly.
	We will be studying symmetric coupling. However, most of the studies can be easily
	extended to asymmetric coupling.
	We give conditions for the stability of the synchronized state
	and demonstrate it with certain examples.
	We propose that the usual definition of Lyapunov exponent using logarithm (which is inverse of exponential) should
	be appropriately modified to obtain an accurate quantifier that
	describes the convergence of trajectories in a stable regime.
	(Mittag-Leffler function is a power-law asymptotically
	which is slower decay than exponential. Thus the
	Lyapunov exponent will
	always be zero if we fit it with an exponential.)
	However,
	the divergence of trajectories in the unstable region is still exponential
	and the usual definition of Lyapunov exponent may hold in this case.

	There are several definitions of fractional differential equations and
	the same is true for fractional difference equations.
	We will study the
	definition obtained by Gejji and Deshpande \cite{gejji}. The evolution depends on
	the value at all previous time-steps. The weight of previous values decays as a power-law and there is a long term memory built-in in the system.
	It is not surprising that the fluctuations also decay extremely slowly
	and decay can be approximated by power-law with power related to the
	order of fractional difference equation. This is in turn related to the
	properties of Mittag-Leffler function which plays an important role
	in fractional differential equations and plays a similar role here.
	In difference equations, the concept of coupled maps was introduced by
	Kapral, Kuznetsov, and Kaneko. Kaneko can be credited for making it
	popular.

	The fractional difference equations have been studied only recently.
	In 1989, Miller and Ross began this investigations\cite{miller}.
	Some of the studies
	in fractional difference equations are
	due to Atici and coworkers \cite{atici1,atici2}, Holm\cite{holm},
	and others
	\cite{wu,mohan}.
	We extend this definition to the coupled fractional difference
	equations. While exploratory works on fractional equivalents
	of known nonlinear maps can yield useful insights, we focus on
	linear systems in this work. Linearizations of nonlinear 
	systems are a standard tool in difference equtions as well as
	coupled difference equations. Most of analytic work in these systems
	is dependent on linearization. Thus understanding linear systems and
	coupled linear systems is extremely important in nonlinear 
	dynamics. We believe that studies in linear systems can be equally
	useful in fractional difference equations.

	\section{Effective Lyapunov Exponent}

	In this section, we introduce an alternative viz. Effective Lyapunov Exponent (ELE) for the classical Lyapunov exponent.
	We will follow the notation
	and definition used by
	Deshpande and Gejji \cite{gejji}.
	They define the fractional equivalent for the 
	$x(n+1)=f(x(n))$ in the following manner.
	They construct an discrete Caputo-type fractional
	difference operator and define
	$u(t)=u_0 + {\frac{1}{\Gamma(\alpha)}} \sum_{j=1}^{t}
	{\frac{\Gamma(t-j+\alpha)} {\Gamma(t-j+1)}}\left( f(j-1,u(j-1))-u(j-1)\right)$
	in general. We assume that 
the function $f$ does not depend on time. We define
	$g_{\alpha}(k)= {\frac{\Gamma(k+\alpha)} {\Gamma(k+1)}}$
	and alternatively write above expression as
	$u(t)=u_0 + {\frac{1}{\Gamma(\alpha)}} \sum_{j=1}^{t}
	g_{\alpha}(t-j) \left(f(u(j-1))-u(j-1)\right)$.

Few one-dimensional maps such as the Gaussian map and Bernoulli map have been
studied in this work. Liu has numerically investigated coupled fractional
Henon map\cite{liu-henon}. 
Henon map is a two-dimensional map and Liu introduces memory in
only one of the variables. We will call such
systems `fractional difference equations of inhomogeneous order'.
	On the other hand, we will investigate maps of homogeneous order.

	%*****************************************

   Consider $f(x)=rx$ where $r\in {\cal R}$
	\begin{equation}
	u(t)=u_0 + {\frac{1}{\Gamma(\alpha)}} \sum_{j=1}^{t}
	g_{\alpha}(t-j) \left(r-1\right)u(j-1)
	\end{equation}
	can be identified with the continuous-time system
	\begin{equation}
	D^\alpha u(t)= (r-1) u(t)\label{b1}
	\end{equation}
	for sufficiently small values of coefficient.
	
\begin{figure}
	\caption{Mittag-Leffler decay.}
	\centering
	\includegraphics[width=0.6\textwidth]{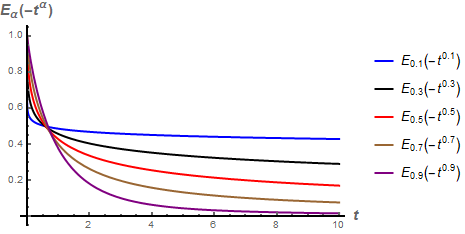}\label{fg0}
	\end{figure}

	The continuous-time system (\ref{b1}) has exact solution in terms of Mittag-Leffler function as below
	\begin{equation}
	u(t)=u(0) E_{\alpha}\left((r-1)t^{\alpha}\right).
	\end{equation}
	Fig. \ref{fg0} shows the Mittag-Leffler decay for various values of $\alpha$.

	This can be alternatively written as
	\begin{equation}
	u(t)=u(0) E_{\alpha}\left(\lambda_e t^{\alpha}\right).
	\end{equation}
	where $\lambda_e$ is effective Lyapunov exponent. For a linear
	first order ordinary differential equation $x'(t)=\lambda x(t)$, the
	solution would be $x(t)=x(0)\exp(\lambda t)$ where
	$\lambda$ is Lyapunov exponent. When it is negative,
	the system goes to absorbing state. Our formulation could be
	considered as generalization

	\begin{equation}
	\lambda_e=\lim_{t\longrightarrow \infty} t^{-\alpha}E_{\alpha}^{-1}\left({\frac{u(t)}{u(0)}}\right).
	\end{equation}

\begin{figure}
	\caption{Computation of $\lambda_e$ for few values of $r$. The value
obtained is $r-1$.}
	\centering
	\includegraphics[width=0.5\textwidth]{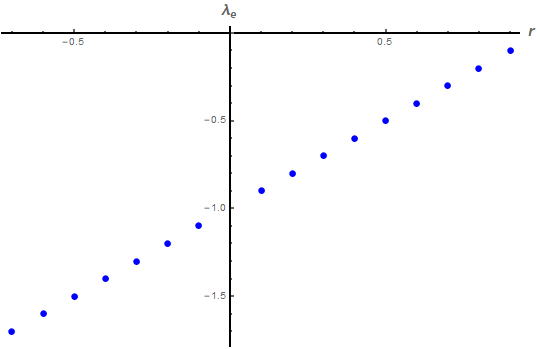} \label{fgn}
	\end{figure}

	This formulation is very similar to standard definition of Lyapunov
	exponent for $\alpha=1$ where $E_{\alpha}(x)=\exp(x)$.
	We will demonstrate that this quantity is a well-defined
	quantity which indeed converges in the stable regime. On the other hand,
if we insist on using the definition of Lyapunov exponent used for
ordinary differential equations, it will lead to zero value. The reason is that
Mittag-Leffler function is a power-law asymptotically which is slower than
	exponential. Like Lyapunov exponent, the effective Lyapunov
	exponent is negative in the absorbing state. In Fig. \ref{fgn}, we sketch the numerically computed $\lambda_e$ for various values of $r$. It is clear that $\lambda_e=r-1$.

	The system is unstable for $r>1$ for any $\alpha$. However, 
	lower bound $a_c(\alpha)$ depends on $\alpha$. Lower bound $a_c(\alpha)
	\rightarrow -1$ as $\alpha\rightarrow 1$. This is an expected 
	limit for integer order difference equation.

	%*******************************************

	\section{Fractional order coupled maps}

	We define two coupled maps in this setting and define.
\begin{eqnarray}
	x(t)&=&x_0 +
{\frac{1}{\Gamma(\alpha)}} \sum_{j=1}^{t} g_{\alpha}(t-j) G(x(j-1),y(t-j)),\nonumber\\
y(t)&=&y_0 +
{\frac{1}{\Gamma(\alpha)}} \sum_{j=1}^{t} g_{\alpha}(t-j) G(y(j-1),x(t-j)) \label{coup}
\end{eqnarray}
	where $G(a,b)=\delta f(a)+q f(b) -a$.\\

	\textbf{Case 1: Real `Normal Modes'}\\

	First we consider the case $f(x)= x$ for which the coefficient matrix of the system has real eigenvalues viz. $\delta+q$ and $\delta-q$.
	Now we consider two new variables $u(t)=x(t)+y(t)$ and $v(t)=x(t)-y(t)$
	and obtain

	\begin{eqnarray}
	u(t)&=&u_0 +
	{\frac{1}{\Gamma(\alpha)}}
	\sum_{j=1}^{t} g_{\alpha}(t-j)
	((\delta+q-1) u(j-1)),\label{a1}\\
	v(t)&=&v_0 +
	{\frac{1}{\Gamma(\alpha)}}
	\sum_{j=1}^{t} g_{\alpha}(t-j)
	((\delta-q-1) v(j-1)).\label{a2}
	\end{eqnarray}

	This is a simple decoupled system of linear difference equations.

	%The variables are decoupled and the equations are

	%practically two independent linear difference equations.

	%The stability region for the above system in $\delta-q$ space is shown

	%in Fig. 1 for various values of fractional order parameter $\alpha$.

	%It is clear that a rhombus constructed from lines $\delta+q=1$,

	%$\delta-q=1$, $\delta+q=a_c(\alpha)$ and $\delta-q=a_c(\alpha)$

	%is the stability region.

	%*****************************************

	Again,
	the discrete-time equations above can be identified with the continuous-time system

	\begin{equation}
	D^\alpha u(t)= (\delta+q-1) u(t)\label{b2}
	\end{equation}

	\begin{equation}
	D^\alpha v(t)= (\delta-q-1) v(t)\label{b3}
	\end{equation}
	for sufficiently small values of coefficient
	and exact solution in terms of Mittag-Leffler function is given as

	\begin{equation}
	u(t)=u(0) E_{\alpha}\left((\delta+q-1)t^{\alpha}\right).
	\end{equation}

	\begin{equation}
	v(t)=v(0) E_{\alpha}\left((\delta-q-1)t^{\alpha}\right).
	\end{equation}

	%*******************************************

	Thus the effective Lyapunov exponents are given by $\delta+q-1$ and
	$\delta-q-1$ for the above decoupled system. We compute the system for
 $T=8 \times 10^4- 10^5$ time-steps. The decay is extremely
slow for smaller $\alpha$. After discarding first 2000 time-steps, we
check 
if the distance of $(x(t),y(t))$ from origin every 100 time-steps and the
trajectory is stable if the distance does not increase till time $T$.

	\begin{figure}
	\caption{Stability regime for symmetrically coupled maps.}
	\centering
	\includegraphics[width=0.5\textwidth]{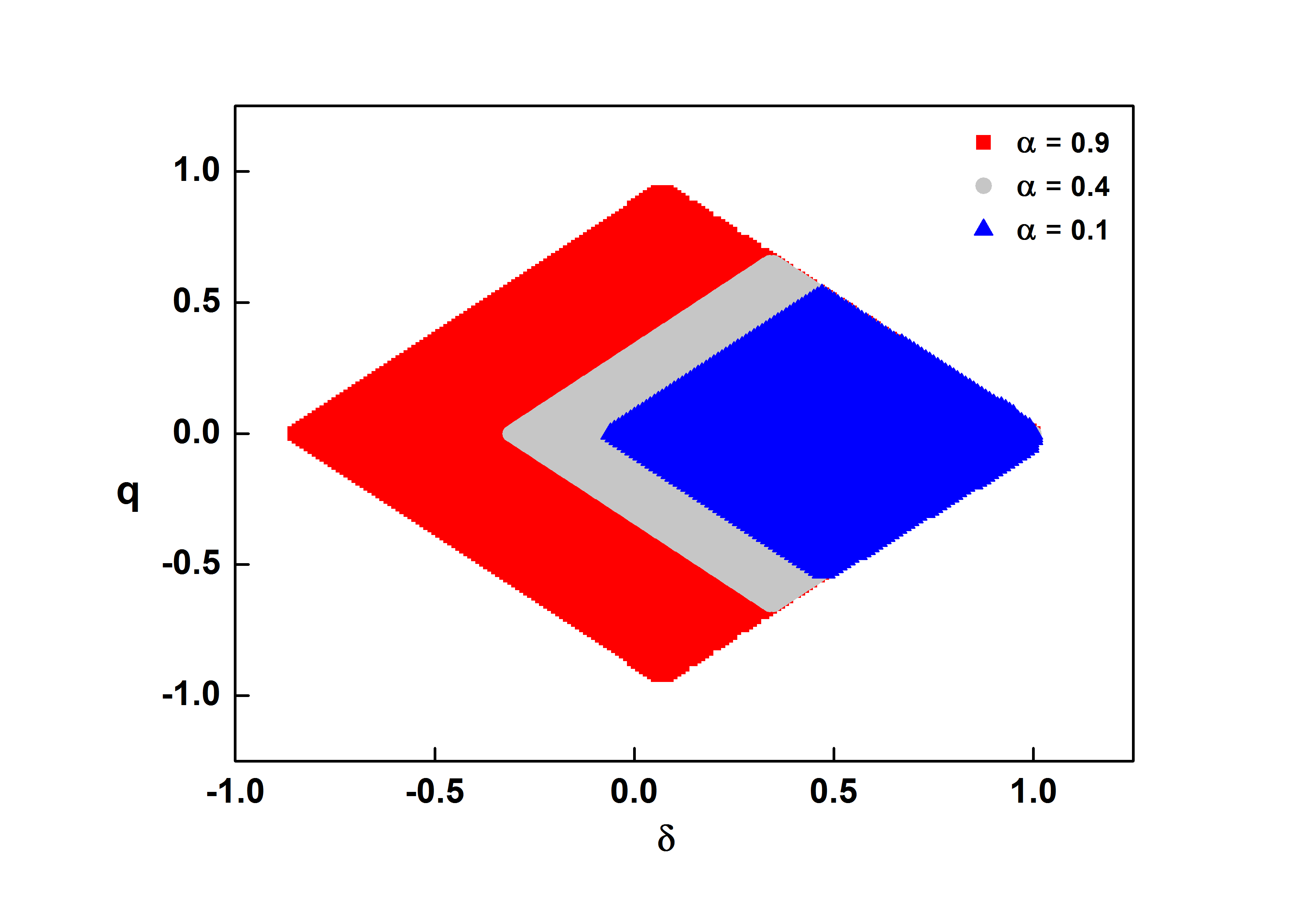} \label{fg3}
	\end{figure}

	The stability regime for the above system is plotted as a function of
	$\delta$ and $q$ in Fig. \ref{fg3}.
	The stability region is a rhombus bounded by lines parallel to
	$\delta=q$ and $\delta=-q$.
	For all values of $\alpha$, the system is unstable for
	$\delta+q-1>0$ and $\delta-q-1>0$. The other two bounding lines
	change with $\alpha$.
	For $q=0$, we have an effectively one-dimensional system.
	The bounds for $q=0$ are $\delta=1$ and $\delta=a_c(\alpha)$.
	As $\alpha \rightarrow 1$, $a_c(\alpha) \rightarrow -1$.
	The lines enclosing stability
	region are given by $\delta \pm q=1$ and $\delta \pm q=a_c(\alpha)$.`
	It is clear from
	$a_c(\alpha)$ are close to $-\alpha$.
	This explains the rhombus structure in Fig. \ref{fg3}. (For $f(x)=rx$, the
phase diagram
will be similar except that the values of $\delta$ and $q$ will be scaled 
to ${\frac{\delta}{r}}$ and ${\frac{q}{r}}$.)\\

\begin{figure}
	\caption{Chaotic synchronization and antisynchronization
		for symmetrically coupled maps.  $\alpha=0.9$,a) $\delta$=0.85,
		$q=-0.2$ and b)$\delta=0.85, q=0.2$}
	\centering
	\includegraphics[width=0.5\textwidth]{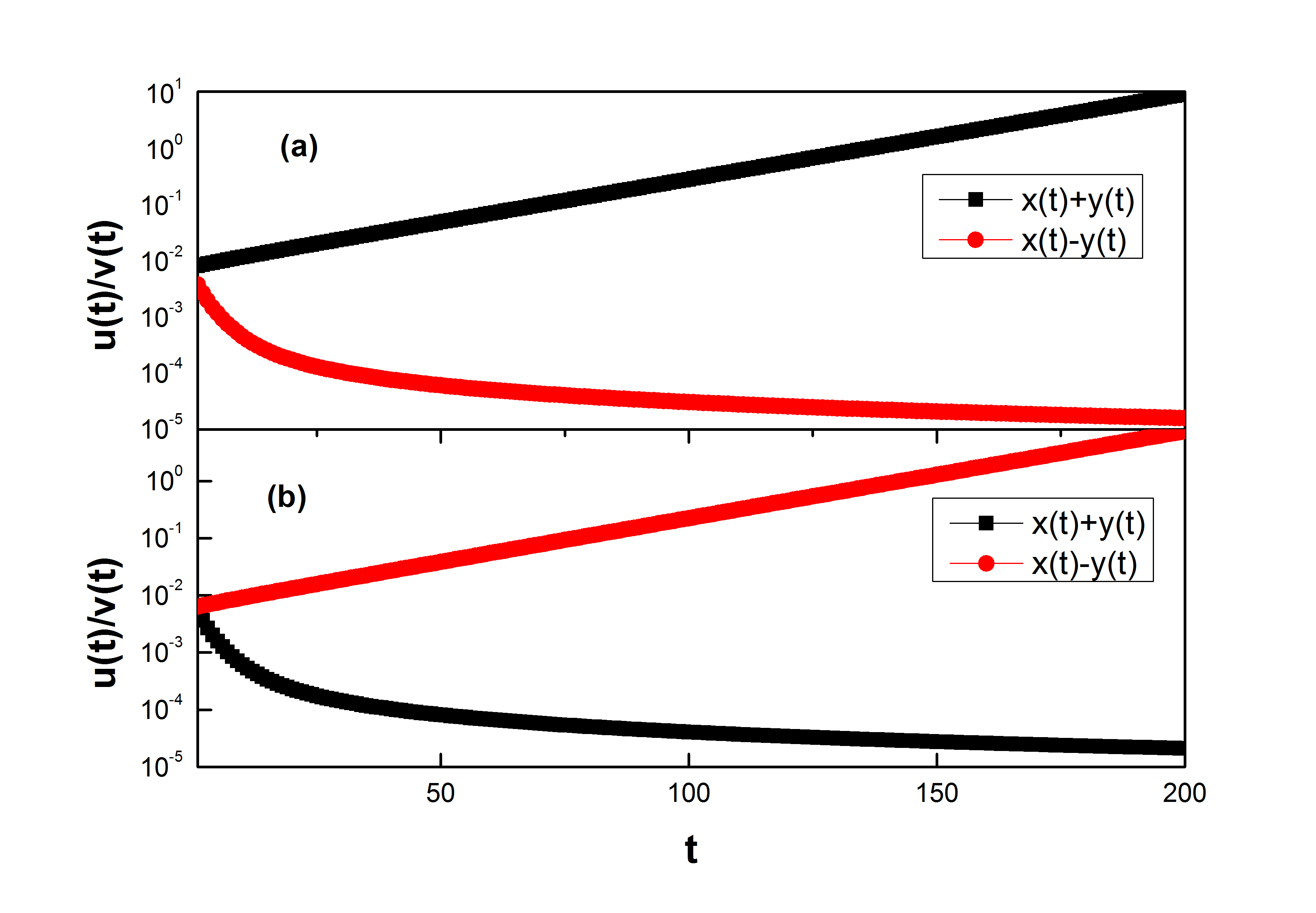} 
	 \label{fg13}
	\end{figure}

	%%%%%%%%%%%%%%%%%%%%%%%%%%%%%%%%%%%%%%%%%%%%%%%
\textbf{Synchronization and antisynchronization}\\
We consider the linear system (\ref{coup}) with real normal modes.
	As shown in (\ref{a1})--(\ref{a2}), 
the sum and difference variables $u(t)=x(t)+y(t)$ and $v(t)=x(t)-y(t)$ are effectively decoupled. 
We can also say the $u$ corresponds to $(1,1)$ mode and $v$ corresonds
to $(1,-1)$ mode.
If the 
	effective Lyapunov exponent corresponding to
	variable $u$ is in the unstable region while the one
	corresponding to $v$ is in stable region, we will
	find that $v(t) \rightarrow 0$. Thus $x(t)\rightarrow y(t)$
implying synchronization. %while the magnitude of $x(t)$ grows. 
On the other hand if $\lambda_e$ corresponding to $v(t)$ is in unstable region and corresponding to 
	$u(t)$ is in the stable region we will observe that $v(t) \rightarrow 0$. This phenomenon is termed as  antisynchronization. Fig. \ref{fg13} shows both these phenomena for different values of parameters.
	We also find that the decaying mode decays slower than exponential
	making it necessary to give a new 
	definition for Lyapunov exponent. However, growing mode
	increases exponentially. The reason may be that approximating
	fractional difference equation by fractional
	differential equation may not be valid for large values.
In coupled map lattices, it is known that the condition for 
chaotic synchronization is that all modes except one corresponding to
the mean, {\it {i.e.}} eigenmode $(1,1,\ldots 1)$ should decay\cite{gade}.
We can obtain  normal modes in fractional system in a similar manner and find
conditions for chaotic synchronization.

	\textbf{Case 2: Complex `Normal Modes'}\\

	Now let us consider another coupled system
	\begin{eqnarray}
	x(t)&=&x_0 +
	{\frac{1}{\Gamma(\alpha)}} \sum_{j=1}^{t} g_{\alpha}(t-j) (\delta x(j-1)+q y(t-j)-x(j-1)),\label{compl1}\\
	y(t)&=&y_0 +
	{\frac{1}{\Gamma(\alpha)}} \sum_{j=1}^{t} g_{\alpha}(t-j) (\delta y(j-1)-q
	x(t-j)-y(j-1)).\label{compl2}
	\end{eqnarray}

	The eigenvalues of coefficient matrix of the system (\ref{compl1})--(\ref{compl2}) are $\delta\pm \iota q$.

	\begin{figure}
	\caption{Stability region for  coupled maps with
antisymmetric coupling. It is clear that the stability region approaches unit circle as $\alpha\rightarrow 1$
while it is  significantly different for small $\alpha$. The stability 
region is shown for different values of $\alpha$ in ascending order in 
a) and in descending order in b). }
	\centering
	\includegraphics[width=0.5\textwidth]{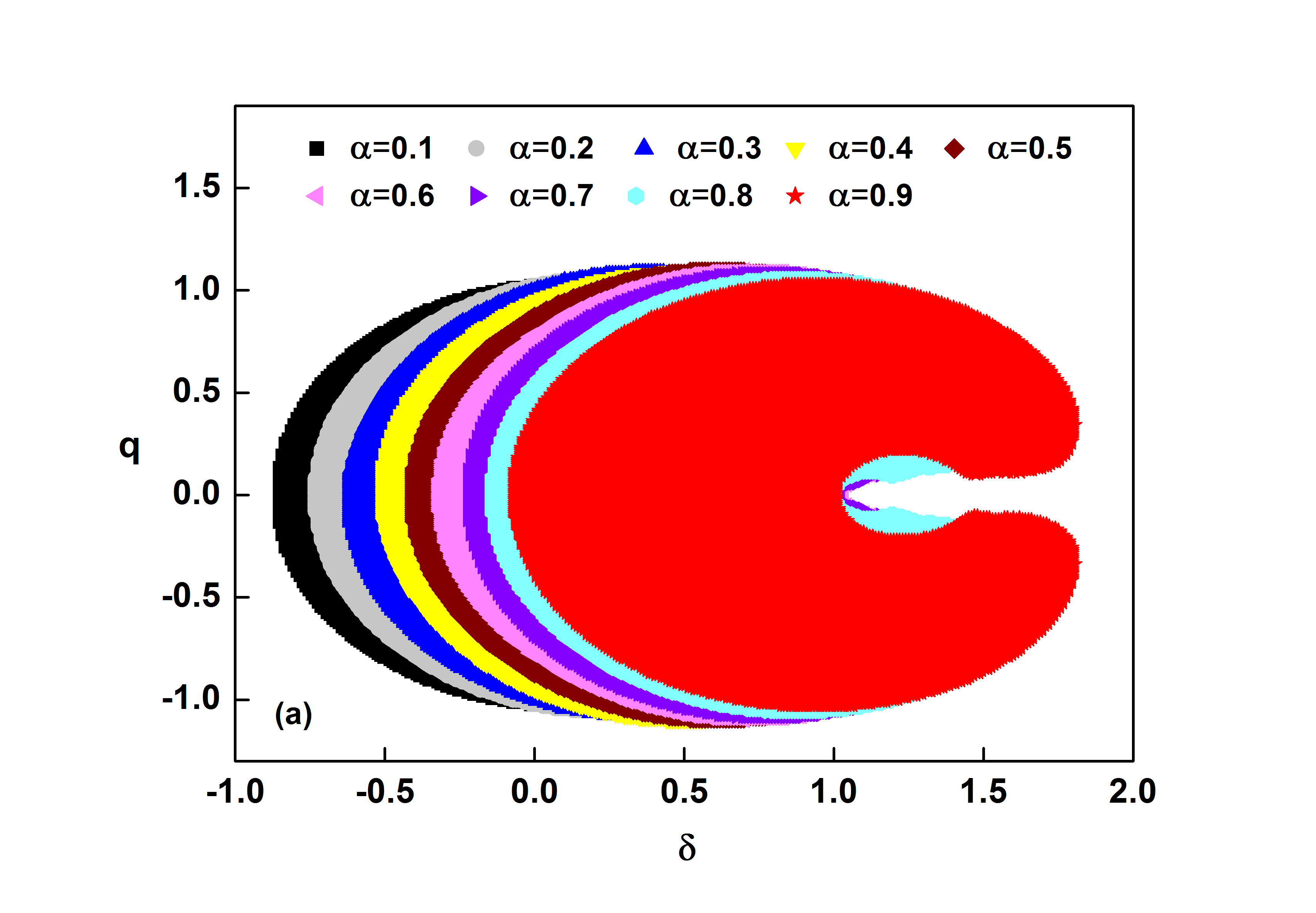}
	\includegraphics[width=0.5\textwidth]{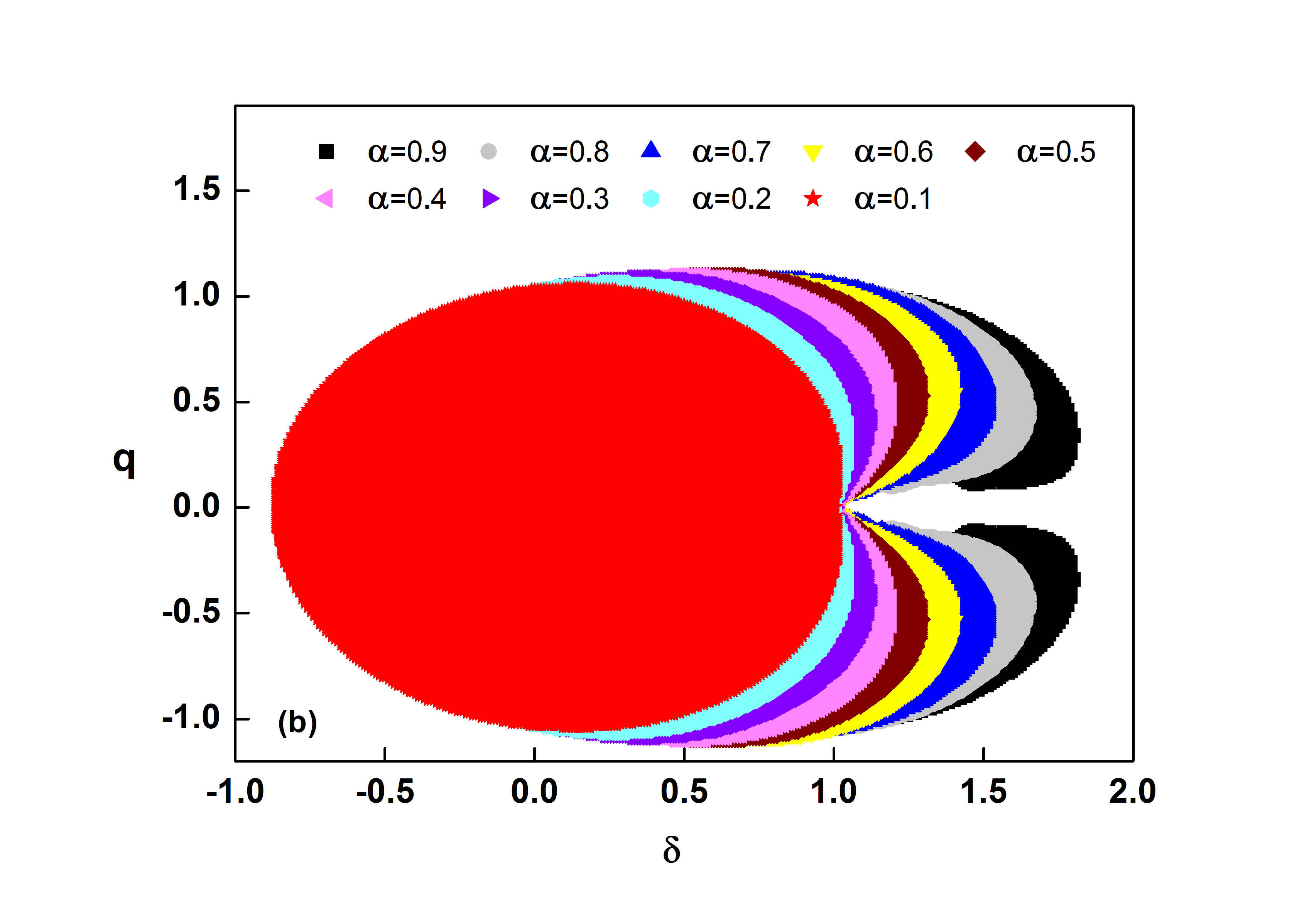}\label{fg4}
	\end{figure}

We have plotted stability region in $\delta-q$ space for various values of
$\alpha$ in Fig. \ref{fg4}. This structure is far richer than one obtained in
Fig. \ref{fg3}. As $\alpha \rightarrow 1$, the stability region tends to the unit
circle in the complex plane which is a stability region for integer-order
difference maps in two dimensions.

	As $\alpha$ decreases, the stability region gradually deforms from an
	unit circle to a non-convex shape.

	Consider $z(t)=x(t)+\iota y(t)$.

	\begin{equation}
	z(t)={\frac{1}{\Gamma(\alpha)}} \sum_{j=1}^{t} g_{\alpha}(t-j)
	\left(((\delta-1) -\iota q) x(j-1)+\iota (\delta-1-\iota q) y(j-1)\right).\\
	\end{equation}

	Thus

	\begin{equation}
	z(t)={\frac{1}{\Gamma(\alpha)}} \sum_{j=1}^{t} g_{\alpha}(t-j)
	(\delta-1) -\iota q) z(j-1).
	\end{equation}

	Similarly, if we define $\bar{z}(t)=x(t)-\iota y(t)$, we get

	\begin{equation}
	\bar{z}(t)={\frac{1}{\Gamma(\alpha)}} \sum_{j=1}^{t} g_{\alpha}(t-j)
	(\delta-1) +\iota q) \bar{z}(j-1).
	\end{equation}

	The variables $z$ and $\bar{z}$ can be associated to ordered
	pairs $(x,y)$ and $(x,-y)$ describing a complex
	number and its conjugate in complex plane.

	%\begin{eqnarray}

	%u(t)&=&{\frac{1}{\Gamma(\alpha)}} \sum_{j=1}^{t} g_{\alpha}(t-j)

	%((\delta-1) +i q) u(j-1)\label{ce1}\\

	%v(t)&=&{\frac{1}{\Gamma(\alpha)}} \sum_{j=1}^{t} g_{\alpha}(t-j)

	%((\delta-1) -i q) v(j-1).\label{ce2}

	%\end{eqnarray}

	%We expect the modulus of these quantities to decay with

	%effective Lyapunov exponent $\sqrt{(\delta-1)^2+q^2}$.

	%We have plotted effective Lyapunov found numerically for various

	%values of $\delta$, $q$ and $\alpha$ as a function of

	%above quantity and it is clear that there is an excellent match.

	The equivalent continuous-time system of (\ref{compl1})--(\ref{compl2}) is given by

	\begin{eqnarray}
	D^\alpha x &=& (\delta-1) x + q y \label{cont1}\\
	D^\alpha y &=& -q x + (\delta-1) y. \label{cont2}
	\end{eqnarray}

	The general solution of system (\ref{cont1})--(\ref{cont2}) is

	\begin{equation}
	x(t)+\iota y(t) = E_\alpha\left((\delta-1-\iota q)t^\alpha\right)\left(x(0)+\iota y(0)\right). \label{cont3}
	\end{equation}

	If we define $z(t)=x(t)+\iota y(t)$ and $\lambda'_e=(\delta-1-\iota q)$, we have

	\begin{equation}
	z(t) = E_\alpha\left(\lambda'_e t^\alpha\right)z(0). \label{cont3}
	\end{equation}

	This motivates us to define the effective Lyapunov exponent $\lambda'_e$
	of the above system as

	\begin{equation}
	\lambda'_e=\lim_{t\longrightarrow \infty}t^{-\alpha}E_{\alpha}^{-1}\left(\frac{z(t)}{z(0)}\right).
	\end{equation}

	This is a well-defined quantity and we always find $\lambda'_e=\delta-1-\iota q$ numerically. This is a complex number. On the other hand, Lyapunov exponents
	for dynamical systems have real Lyapunov exponents. We believe that
	a single real number can determine stability for difference equation because
	the stability condition for integer-order differential equation
	is that the real part of eigenvalues is less than zero. This is a
	single condition. On the other, for fractional order differential
	equations, stability condition is given by two lines. Thus a single
	number may not be enough to determine stability of fractional
	equations. However, if we insist that
	the effective Lyapunov exponents should be real, we can take
	$-\vert \lambda'_e\vert$ as an effective Lyapunov exponent.
	Alternatively,
	we may define the effective Lyapunov exponent $\lambda_e$ of system (\ref{cont1})--(\ref{cont2}) as

	\begin{equation}
	\lambda_e=\lim_{t\longrightarrow \infty}\left|t^{-\alpha}E_{\alpha}^{-1}\left(\frac{x(t)+\iota y(t)}{x(0)+\iota y(0)}\right) \right|=\sqrt{(\delta-1)^2+q^2}.
	\end{equation}

	%If $(\delta, q)$ belongs to the stability region then the absolute values of the quantities $u(t)=ix(t)+y(t)$ and $v(t)=-ix(t)+y(t)$ decay as $\vert E_\alpha\left((\delta-1-\iota q)t^\alpha\right) \vert$.\\

	Note that, $\frac{\sqrt{x(t)^2+y(t)^2}}{\sqrt{x(0)^2+y(0)^2}}=\vert E_\alpha\left((\delta-1-\iota q)t^\alpha\right) \vert$. In general, it is not possible to find the inverse of composite function on right side. However, for $\alpha\in [0.5, 1]$ and $(\delta, q)$ in the stability region we observed that the quantity $\frac{\sqrt{x(t)^2+y(t)^2}}{\sqrt{x(0)^2+y(0)^2}}$ decays with $\lambda_e$ and we have $\lambda_e\approx \lim_{t\longrightarrow \infty}t^{-\alpha}E_{\alpha}^{-1}\left(\frac{\sqrt{x(t)^2+y(t)^2}}{\sqrt{x(0)^2+y(0)^2}}\right)$.

\begin{figure}

	\caption{Effective Lyapunov exponent $\lambda_e$ is
	plotted as function of expression
	$\sqrt{(\delta-1)^2+q^2}$.}
	\centering
	\includegraphics[width=0.6\textwidth]{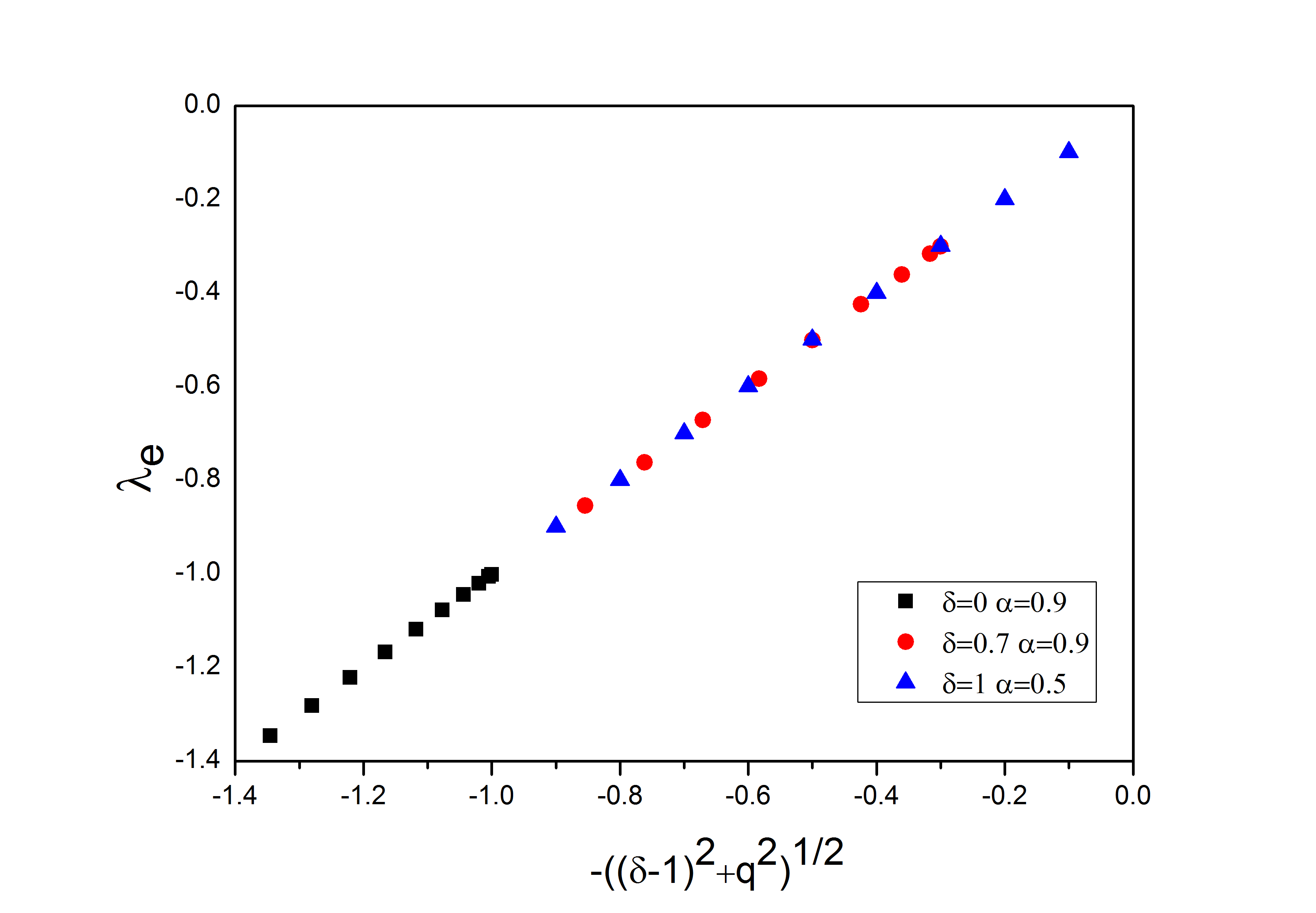} \label{fg5}
	\end{figure}

	In Fig. \ref{fg5} we have plotted effective Lyapunov found numerically for various
	values of $\delta$, $q$ and $\alpha$ as a function of
	above quantity and it is clear that there is an excellent match.

\section{Generalization to nonlinear case}
	Let us try to extend the analysis to nonlinear systems. Let us consider
	stability of the fixed point of a nonlinear system.
Nonlinear system can be linearized by $f(x)=f(x^*)+f'(x^*)(x-x^*)+\ldots$.
	We define $y=x-x^*$. Now $x_{n+1}=f(x_n)$ is equivalent to
	$y_{n+1}=x_{n+1}-x^*=f(x_n)-x^*=f'(x^*)y_n=ry_n$ where
	$r=f'(x^*)$.
	Thus we can conjecture
	that the stability regime for the fixed point is
	given by $a_c(\alpha)<f'(x^*)<1$.
	Interestingly, this conjecture works.
	Consider the cases
	of Bernoulli map 
	and Gauss map 
	considered by Deshpande and Gejji \cite{gejji}.

\begin{figure}

	\caption{a)Bifurcation diagram for Gauss map
	as a function of $\beta$ for various values of $\alpha$. b) Local slope of fixed point $x^*$ is plotted as a function of
	$\beta$. Vertical lines show the values of $\beta$ at which
	$f'(x^*)=a_c(\alpha)$ for these values of $\alpha$.
	The fixed point is stable in the expected region even for a
	nonlinear function.}
	\centering
	\includegraphics[width=0.6\textwidth]{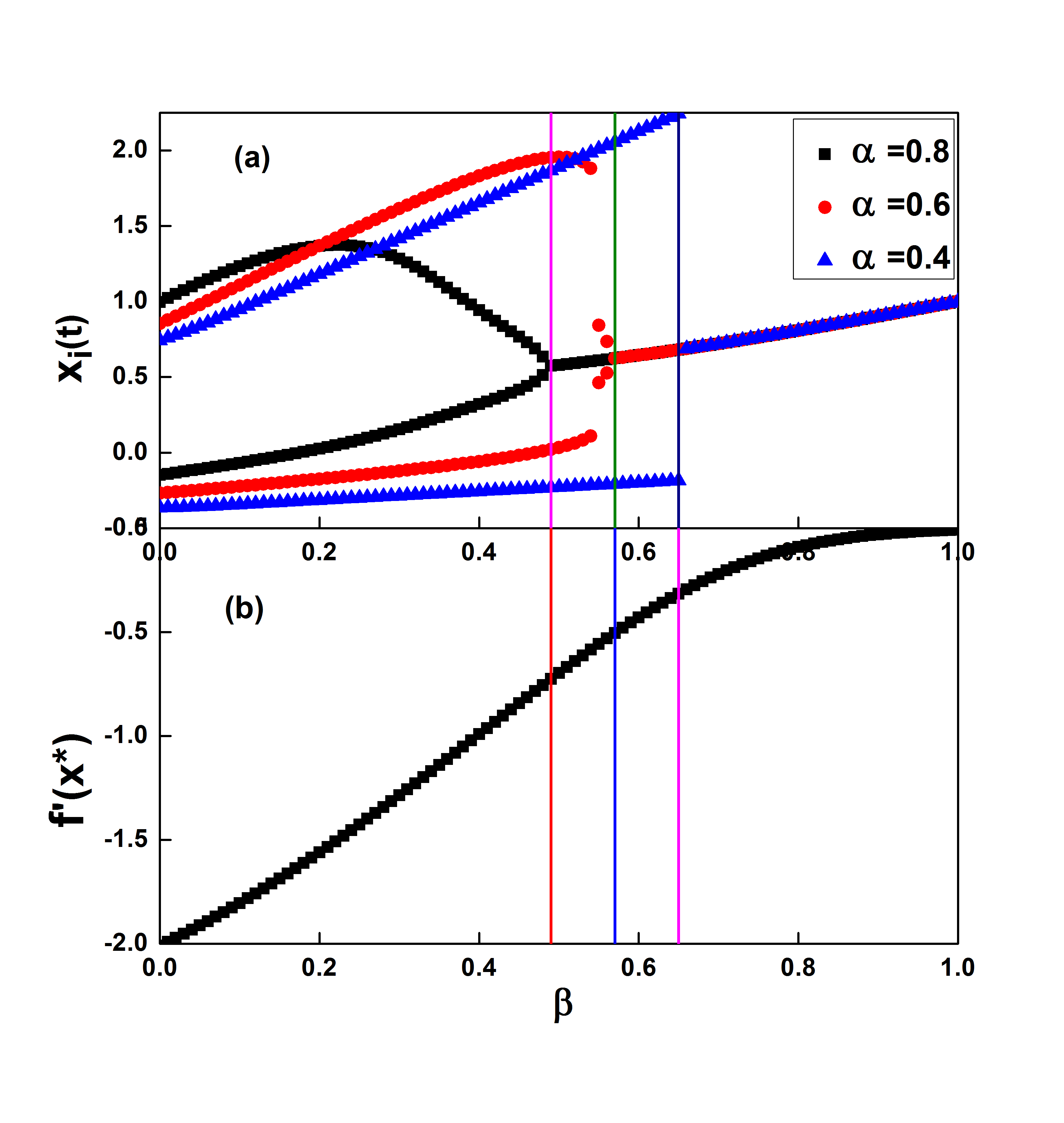} \label{fg6}
	\end{figure}

For Bernouli map
$f(x)=r x\vert_{mod 1}, r>0$, we can guess that for $r<1$, the fixed point is stable
for any value of $\alpha$.
	This is precisely what Deshpande and Gejji \cite{gejji} find.
	In the
	case of Gauss map given by
$f(x)=\exp(-7.5 x^2)+\beta$, the fixed point can be found numerically 
using bisection method or other methods. The fixed point is close
	to $\beta$. The slope is negative and the stability regime is
	dependent on value of $\alpha$. One expects this fixed point to
	be more stable if $\alpha$ increases. It can be 
checked that $a_c(0.4)=-0.3132$, $a_c(0.6)=-0.5085$
	and $a_c(0.8)=-0.7328$. The critical points
corespond to values of $\beta$ at which the $f'(x^*)$ matches with
these values. Thus we expect the fixed point to
	be stable for $\beta>0.49$ for $\alpha=0.8$, for $\beta>0.57$ for $\alpha=0.6$
	and $\beta >0.65$ for $\alpha=0.4$ (cf. Fig. \ref{fg6}).

This can be confirmed from the above paper \cite{gejji} as well as independent simulations.
	We start with $\beta$ as an initial condition.

	\section{Discussion and Conclusions}

Analytic studies in nonlinear dynamics  or coupled nonlinear 
systems are often based on local linearization of dynamics. 
Thus linear systems serve as a basis which helps undertanding
dynamics in these systems. We have studied the fractional equivalent
of linear maps, which are not studied before
to our knowledge. Our studies indicate that the results obtained
are useful in studies of  nonlinear systems as well.
For coupled systems, we have studied the stability of the fixed point.
We also find that the conditions for chaotic synchronization and
antisynchronization and find that the conditions are very
similar to those obtained for coupled integer order maps.

	For one-dimensional $f(x)=r x$, we find that
	the stability regime is given by $a_c(\alpha)<r<1$.
	In the stable regime the dynamics is governed by Mittag-Leffler 
	function. We also define the effective Lyapunov exponent and find
	that $r-1$ is effective Lyapunov exponent in this case. 

	For two coupled linear systems, the behavior
is different for symmetric and antisymmetric coupling.
 The analysis is
	motivated by study of linear difference equations which is 
	essentially the theory of matrices. We can reduce dynamics to 
	`normal modes' which could be real or complex. We can find effective
	Lyapunov exponents in these cases as well. When the normal modes 
	are real, the stability condition is the same as the condition for a
	single linear map for each mode. A much richer picture  is observed
	for complex normal modes. The effective Lyapunov exponents, in this
	case, are complex which is not entirely unexpected.

The stability region of a continuous-time fractional-order dynamical 
system is a superset of that of classical integer-order one. As we 
increase the fractional order to 1, the cone-like stability region of 
the fractional case gets contracted to the left-half complex plane and 
we get the usual region of stability of classical case. In this 
article, we showed that the stability properties of discrete-time 
systems are different. The cardioid-like stability region of 
fractional order system gets deformed and converted to the unit disc 
as we increase the order to 1.

	We also extend this work to fixed points of nonlinear maps and
	confirm that a  similar criterion holds.
This work can be extended in many directions. We can try to find the
stability of periodic orbits of the higher period and find routes to chaos
in low-dimensional fractional difference equations.

	\section*{Acknowledgements}

	PMG thanks DST-SERB for financial assistance (Ref. EMR/2016/006686). SBB acknowledges the Science and Engineering Research Board (SERB), New Delhi, India for the Research Grant (Ref. MTR/2017/000068) under Mathematical Research Impact Centric Support (MATRICS) Scheme.

	\bibliography{mybib.bib}

\begin{thebibliography}{10}

\bibitem{may}
R.~M. May, ``Biological populations obeying difference equations: stable
  points, stable cycles, and chaos,'' {\em Journal of Theoretical Biology},
  vol.~51, no.~2, pp.~511--524, 1975.

\bibitem{strogatz}
S.~H. Strogatz, {\em Nonlinear dynamics and chaos with student solutions
  manual: With applications to physics, biology, chemistry, and engineering}.
\newblock CRC press, 2018.

\bibitem{feigenbaum1}
M.~J. Feigenbaum, ``The universal metric properties of nonlinear
  transformations,'' {\em Journal of Statistical Physics}, vol.~21, no.~6,
  pp.~669--706, 1979.

\bibitem{feigenbaum2}
M.~J. Feigenbaum, ``Quantitative universality for a class of nonlinear
  transformations,'' {\em Journal of statistical physics}, vol.~19, no.~1,
  pp.~25--52, 1978.

\bibitem{ott}
E.~Ott, {\em Chaos in dynamical systems}.
\newblock Cambridge university press, 2002.

\bibitem{pakhare}
S.~S. Pakhare, V.~Daftardar-Gejji, D.~S. Badwaik, A.~Deshpande, and P.~M. Gade,
  ``Emergence of order in dynamical phases in coupled fractional gauss map,''
  {\em Chaos, Solitons \& Fractals}, vol.~135, p.~109770, 2020.

\bibitem{gejji}
A.~Deshpande and V.~Daftardar-Gejji, ``Chaos in discrete fractional difference
  equations,'' {\em Pramana}, vol.~87, no.~4, p.~49, 2016.

\bibitem{miller}
K.~S. Miller and B.~Ross, ``Fractional difference calculus.,'' in {\em
  Proceedings of the International Symposium on Univalent Functions, Fractional
  Calculus and Their Applications, Nihon University, Koriyama, Japan,}, p.~139,
  1989.

\bibitem{atici1}
F.~M. At{\i}c{\i} and S.~{\c{S}}eng{\"u}l, ``Modeling with fractional
  difference equations,'' {\em Journal of Mathematical Analysis and
  Applications}, vol.~369, no.~1, pp.~1--9, 2010.

\bibitem{atici2}
F.~M. Atici and P.~W. Eloe, ``A transform method in discrete fractional
  calculus,'' {\em International Journal of Difference Equations}, vol.~2,
  no.~2, 2007.

\bibitem{holm}
M.~T. Holm, ``The theory of discrete fractional calculus: Development and
  application,'' 2011.

\bibitem{wu}
G.-C. Wu and D.~Baleanu, ``Chaos synchronization of the discrete fractional
  logistic map,'' {\em Signal processing}, vol.~102, pp.~96--99, 2014.

\bibitem{mohan}
J.~J. Mohan and G.~Deekshitulu, ``Fractional order difference equations,'' {\em
  International Journal of Differential Equations}, vol.~2012, 2012.

\bibitem{liu-henon}
Y.~Liu, ``Discrete chaos in fractional hi{\'e}non maps,'' {\em International
  Journal of Nonlinear Science}, vol.~2018, pp.~170--175, 2014.

\bibitem{gade}
P.~M. Gade, ``Synchronization of oscillators with random nonlocal
  connectivity,'' {\em Physical Review E}, vol.~54, no.~1, p.~64, 1996.

\end{thebibliography}

	\bibliographystyle{ieeetr}

	\end{document}